\documentclass[aps,prl,superscriptaddress,amsmath,amssymb,twocolumn,a4paper,floatfix]{revtex4-2}
\usepackage[utf8]{inputenc}
\usepackage[english]{babel}
\usepackage{graphicx}
\usepackage{hyperref}
\usepackage{natbib}
\usepackage{float}
\usepackage{xcolor}
\usepackage{bbm}

\DeclareUnicodeCharacter{2212}{-}

\begin{document}
\title{Strange Metal Solution in the Diagrammatic Theory for the $2d$ Hubbard Model}

\author{Aaram J. Kim}
\affiliation{Department of Physics, King's College London, Strand, London WC2R 2LS, UK}
\affiliation{Department of Physics, University of Fribourg, Chemin du Mus\'ee 3, 1700 Fribourg, Switzerland}

\author{Philipp Werner} 
\affiliation{Department of Physics, University of Fribourg, Chemin du Mus\'ee 3, 1700 Fribourg, Switzerland}

\author{Evgeny Kozik}
\affiliation{Department of Physics, King's College London, Strand, London WC2R 2LS, UK}

\begin{abstract}
We show that the numerically exact bold-line diagrammatic theory for the $2d$ Hubbard model exhibits a non-Fermi-liquid (NFL) strange metal state, which is connected to the SYK NFL in the strong-interaction limit. 
The solution for the doped system features the expected phenomenology with the NFL near half-filling at strong couplings and in a wide temperature range enclosed by the atomic state at high temperatures and a Fermi liquid at low temperatures. We demonstrate, however, that this behavior in the weakly doped regime is due to the unphysical branch of the Luttinger-Ward functional. On the other hand, our analysis shows that the NFL physics is realized at larger doping.
\end{abstract}

\maketitle
The strange metal in unconventional superconductors~\cite{Stewart2001,Stewart2006,Kasahara2010,Dressel2011,Keimer2015} is one of the most intriguing phenomena in condensed matter physics.
Universal experimental signatures demonstrate clear deviations from conventional Fermi-liquid (FL) theory: temperature ($T$)-linear in-plane resistivity and violation of the Mott-Ioffe-Regel (MIR) limit~\cite{Takagi1992,Martin1990,Valla2000}, robust power-law behavior in the optical conductivity~\cite{ElAzrak1994,Marel2003,Basov2005}, $T$-linear magnetoresistivity~\cite{Hayes2016,Giraldo-Gallo2018}, and strong temperature dependence in the NMR response in contrast to Korringa's law~\cite{Berthier1994}.
The $2d$ Hubbard model~\cite{Hubbard} 
is widely believed to incorporate the rich physics of high-$T_c$ superconductivity, including the strange metal phase~\cite{Scalapino2012,Brown2019}.
The basic model Hamiltonian is 
\begin{equation}
	\mathcal{H}_{\text{HM}} = -t\sum^{}_{\langle ij\rangle\sigma}\hat{c}^{\dagger}_{i\sigma}\hat{c}^{}_{j\sigma} -\mu\sum^{}_{i\sigma}\hat{n}_{i\sigma} + U\sum^{}_{i}\hat{n}_{i\uparrow}\hat{n}_{i\downarrow}~,
	\label{eqn:Hhm}
\end{equation}
where $t$, $\mu$, and $U$ represent the nearest-neighbor hopping amplitude, chemical potential, and on-site repulsion, respectively.
$\hat{c}^{}_{i\sigma}$ ($\hat{c}^{\dagger}_{i\sigma}$) is the annihilation (creation) operator of a spin-$\sigma$ fermion at site $i$ and $\hat{n}_{i\sigma} = \hat{c}^{\dagger}_{i\sigma}\hat{c}^{}_{i\sigma}$~.

An alternative picture of the strange metal phase 
is provided by the SYK model~\cite{Sachdev:1993hr,Parcollet1999,Georges2001,Kitaev:2015,Gu2020}. 
Originally introduced 
to describe 
quantum spins with all-to-all interactions and quenched disorder, the model in complex fermionic representation with combined spin and site indices can be written as
\begin{equation}
	\mathcal{H}_{\text{SYK}} = \sum^{N}_{abcd}J_{abcd}\hat{c}^{\dagger}_{a}\hat{c}^{\dagger}_{b}\hat{c}^{}_{c}\hat{c}^{}_{d} - \mu\sum^{N}_{a}\hat{n}_a~.
	\label{eqn:Hsyk}
\end{equation}
Here, $J_{abcd}$ are the random couplings between the fermionic degrees of freedom, which follow a Gaussian distribution with zero mean and $U^2$ variance. The model is exactly solvable in the large-$N$ and low-$T$ limit, and its solution of a specific NFL form was shown to reproduce the strange metal phenomenology~\cite{Varma1989,Parcollet1999} of high-$T_c$ superconductors 
in its lattice extensions~\cite{Song2017,Patel2018,Chowdhury2018}. Despite the different microscopic ingredients of models (\ref{eqn:Hhm}) and (\ref{eqn:Hsyk}), their 
strikingly similar strange metal properties make the possibility of underlying connections between them a topic of intensive research~\cite{Werner2018,Tsuji2019,Cha2019,Cha2020}. 

One such connection is of mathematical nature, expressed by the diagrammatic correspondence between the SYK model and the $t \ll T,U$ limit of the Hubbard model, the isolated Hubbard atom (HA)~\cite{Cha2019}. After averaging over the Gaussian disorder, the saddle point equations of the SYK model take the form~\cite{Sachdev:1993hr}
\begin{align}
	G(i\omega_n) &= \frac{1}{i\omega_n + \mu - \Sigma(i\omega_n)}~,
	\Sigma(\tau) = -U^2G(\tau)^2G(-\tau)~,
	\label{eqn:sc}
\end{align}
where $G$ and $\Sigma$ represent the interacting 
Green's function and self-energy in imaginary-time $\tau$ or Matsubara frequency $\omega_n$. On the other hand, Eq.~(\ref{eqn:sc}) is exactly the self-consistency condition of the bold-line (skeleton) diagrammatic theory for the Green's function of the HA, truncated at the second order in $U$. In regimes where the truncation error is small, this formal correspondence suggests a solution of the Hubbard model that features a strange metal NFL behavior with properties similar to those predicted by the SYK model. This could be a scenario that connects the phenomenologies of the different microscopic models, and further supports the status of the Hubbard model as the ``standard model'' for high-$T_c$ superconductivity. Testing it appears straightforward, since the skeleton diagrammatic series can be evaluated to high orders without approximations~\cite{Prokofev2007,Prokofev2008,VanHoucke2012,Davody2013,Kulagin2013a,Kulagin2013,Mishchenko2014,Deng2015,Huang2016,Simkovic2017}. 

However, this argument has a potential loophole: Convergence of the expansion in $U$ and the self-consistently determined Green's function to a unique solution does not guarantee that this solution is physical. The underlying Luttinger-Ward functional (LWF) has multiple branches~\cite{Schafer2013,Kozik2015,Stan:2015en,Rossi:2015cx,Gunnarsson:2017co,Tarantino:2017dra,Chalupa2018,Reitner2020,Kim2020}, which in certain regimes is known to result in convergence of the bold-line series to an unphysical solution~\cite{Kozik2015}.
Therefore, even if a SYK-like NFL diagrammatic solution is found for the Hubbard model, it needs to be scrutinized to prove its validity. 

Here we use the numerically exact bold-line diagrammatic Monte Carlo (DiagMC) method~\cite{Prokofev2007,Prokofev2008,VanHoucke2012,Davody2013,Kulagin2013a,Kulagin2013,Mishchenko2014,Deng2015,Huang2016,Simkovic2017}, to obtain the diagrammatic solution of the Hubbard model [Eq.~(\ref{eqn:Hhm})] in the strongly correlated regime. In this approach, all Feynman diagrams for the expansion in powers of $U$ of the self-energy $\Sigma[G]$ in terms of the full Green's function $G$ are summed stochastically to high order until convergence; then a new estimate for $G=G[\Sigma]$ is obtained from the Dyson equation, and the scheme is iterated until self-consistency. At intermediate $T$ and for a range of doping we obtain a NFL solution with local spectral properties resembling those of the SYK model [Eq.~(\ref{eqn:Hsyk})]. 
According to Ref.~\cite{Rossi_shifted_action_2016}, the solution $G$ is physical if the resulting diagrammatic series for $\Sigma[G]$ is inside its convergence radius. This implies that, in order to yield an unphysical result, the self-consistency condition must tune the series precisely to its convergence radius.  
We find this to be the case for the NFL solutions in the underdoped regime. 
These solutions are, therefore, remarkable examples of unphysical results produced by the diagrammatic theory that appear to be physically sound, highly non-trivial, and consistent with general expectations. Further analysis of the series properties however indicates that the NFL solution becomes physical in the overdoped regime. 

\begin{figure}[t]
	\centering
	\includegraphics[width=0.45\textwidth]{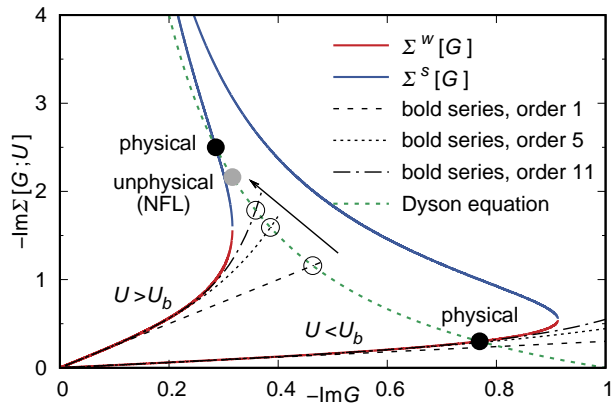}
	\caption{
		Mechanism of the unphysical solution (of SYK-type NFL character in the Hubbard atom and the intermediate-$T$ $2d$ Hubbard model near half-filling) in the bold-line diagrammatic theory, illustrated schematically by the solution of the $(0+0)d$ toy model~\cite{Rossi:2015cx,Felix2016, Kim2020}. The functional $\Sigma[G]$ has two branches $\Sigma^w$ (red) and $\Sigma^s$ (blue). The partial sums $\Sigma^{(n)}[G]$ for $n=1,5,11$ of the diagrammatic expansion of $\Sigma^w[G]$ are the black dashed, dotted, and dash-dotted lines, respectively. A self-consistent solution satisfies the Dyson equation (green dashed line). 
	}
	\label{fig:0dmodel}
\end{figure}

Feynman diagrams are a common tool in quantum many-body physics~\cite{AGD}. While bare expansions in terms of the non-interacting Green's function $G_0$ rely on the 
arbitrary 
partition of the Hamiltonian into free particles and their interaction, the bold-line technique offers a physically appealing picture: (i) $G$ is measurable, (ii) as revealed by Baym and Kadanoff~\cite{Baym_Kadanoff}, expansions in terms of $G$ automatically respect conservation laws, (iii) they contain fewer diagrams since $G$ itself represents an infinite diagrammatic series, which is the essence of renormalizations that eliminate unphysical divergences. Ironically, divergences are manageable (by resummation methods), while the convergence of bold-$G$ series to the unphysical answer~\cite{Kozik2015} is fatal. 

In the bold-line technique relevant for Eq.~\eqref{eqn:sc}~\cite{AGD}, $G$ is determined by consecutive approximations $G^{(n)}$, $G=\lim_{n \to \infty} G^{(n)}$, that solve the Dyson equation 
\begin{equation}
[G^{(n)}]^{-1}=G_0^{-1} - \Sigma^{(n)}[G^{(n)}]. \label{eqn:Dyson}
\end{equation}
Here the functional $\Sigma^{(n)}[G]$ is the sum up to order $n$ of the bold-line expansion in powers of $U$: 
\begin{equation}
\Sigma^{(n)}[G]=\sum_{m=1}^{n} a_m[G] (\xi U)^m, \label{eqn:series}
\end{equation}
where the series coefficients $a_m[G]$ depend only on the function $G$ and $\xi$ is a formal expansion parameter set to $\xi=1$ in final expressions. Misleading convergence results from the existence of at least two branches of the LWF for the Hubbard interaction (actually, infinitely many branches can be found~\cite{Gunnarsson:2017co}), as illustrated schematically in Fig.~\ref{fig:0dmodel}, which shows the solution of the $(0+0)$-dimensional toy model~\cite{Rossi:2015cx,Felix2016,Kim2020}.
For arbitrary $G$, the functional $\Sigma[G]$ consists of the ``weakly-correlated'' branch $\Sigma^w[G]$ (red line) and the ``strongly-correlated'' branch $\Sigma^s[G]$ (blue line), depicted in Fig.~\ref{fig:0dmodel} at two characteristic values of $U$. 
Being constrained by the Dyson equation, the physical solution $G_{\mathrm{ph}}, \Sigma_{\mathrm{ph}}$ lies at the intersection of $G_0^{-1}-G^{-1}$ (the dashed green line) and $\Sigma[G]$. 
The key observation is that $\Sigma_{\mathrm{ph}}$ switches from $\Sigma^w[G]$ at $U<U_b$ to $\Sigma^s[G]$ at $U>U_b$ ($U_b$ is the interaction at which the solution is precisely at the branching point), while $\Sigma^{(n)}[G]$ is the power series only for $\Sigma^w[G]$ by construction. 
Therefore, at $U>U_b$, the sequence $G^{(n)}$ cannot converge to $G_{\mathrm{ph}}$. 
Although 
for each fixed $n$ a solution of Eq.~(\ref{eqn:Dyson}) 
can be found (open circles), the sequence $G^{(n)}$ approaches an unphysical fixed point (gray circle), which is located precisely at the boundary of convergence of the series $\Sigma^{(n)}[G]$ as a functional of $G$. 

\begin{figure}[t]
	\centering
	\includegraphics[width=0.45\textwidth]{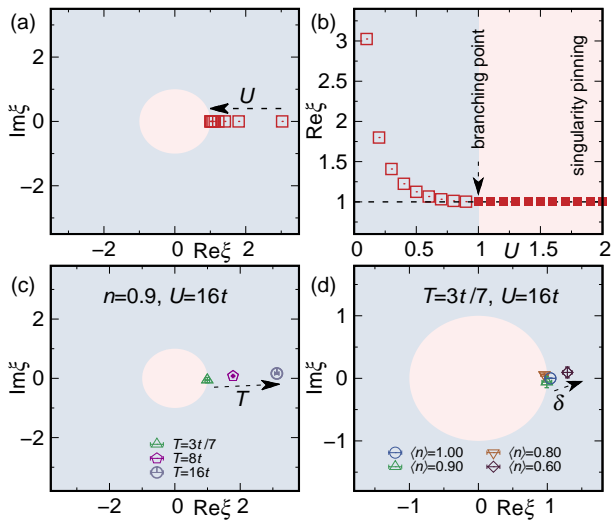}
	\caption{
		Singularity location $\xi_s$ of the diagrammatic solution (\ref{eqn:Dyson}), (\ref{eqn:series}) in the $n \to \infty$ limit for the $(0+0)d$ toy model at various $U$ [(a), (b)] and for the $2d$ Hubbard model at $U=16t$ for various temperatures [at $10\%$ doping, (c)] and doping $\delta$ [at $T=3t/7$, (d)].
	}
	\label{fig:singularitymap}
\end{figure}

This mechanism allows one to formulate~\cite{Rossi_shifted_action_2016} a practical criterion for detecting unphysical solutions of the diagrammatic technique [Eq.~(\ref{eqn:Dyson}),~(\ref{eqn:series})]. If for the final result, $G=\lim_{n \to \infty} G^{(n)}$, the series $\Sigma^{(n)}[G]$ is strictly within its convergence radius, then $\lim_{n \to \infty} \Sigma^{(n)}[G] = \Sigma^w[G]$ and $G$ is necessarily the correct physical solution. If, on the contrary, $\Sigma^{(n)}[G]$ happens to be at its convergence radius, $G$ is unphysical, possibly except for a space of Hamiltonian parameters of measure zero. For a fixed $G$, the convergence radius is given by $|\xi_s|$, $\xi_s$ being the location of the singularity nearest to the origin in the complex plane of $\xi$ extracted from the coefficients $a_n[G]$. Therefore, an unphysical solution is signalled by \textit{pinning} of the singularity at $|\xi_s|=1$ while the Hamiltonian parameters are varied in a non-trivial range. This effect is illustrated for the exactly-solvable $(0+0)$-dimensional model in the top panels of Fig.~\ref{fig:singularitymap}, which show $\xi_s$ as a function of $U$. At small $U<U_b$, where the diagrammatic solution is physical, $\mathrm{Re} \xi_s >1$, but it moves towards the origin as $U$ is increased, reaching $\xi_s=1$ at $U=U_b$ and remaining pinned there for all $U>U_b$.


\begin{figure}[t]
	\centering
	\includegraphics[width=0.45\textwidth]{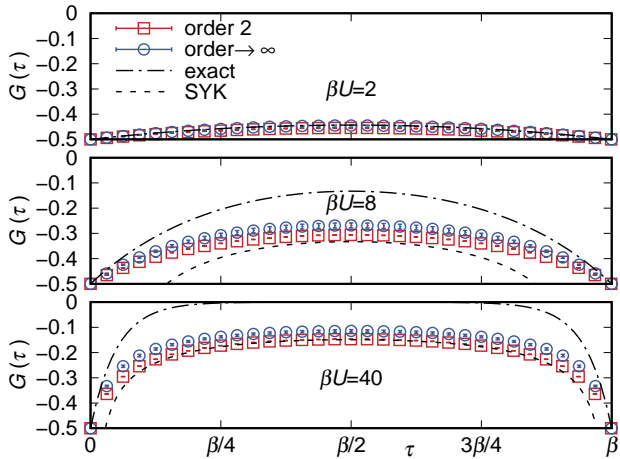}
	\caption{ 
	          Bold-line diagrammatic solution $G(\tau)$ of the half-filled HA at various $\beta U$ truncated at second order ($n=2$, squares) and converged with the diagram order (circles). The exact solution of the HA and the long-time asymptotics of the saddle-point solution of the SYK model are the dash-dotted and dashed line, respectively.
	}
	\label{fig:scGt}
\end{figure}

Central to the 
nature of the many-body state is the long-time asymptotic scaling of the local Green's function 
\begin{equation}
	\log\left[-G_{\mathrm{loc}}(\tau)\right] = C-2\varepsilon \left( \tfrac{\pi\delta\tau}{\beta}\right) + \Delta \left(\tfrac{\pi\delta\tau}{\beta}\right)^2~,
	\label{eqn:Gansatz}
\end{equation}
with $[\delta\tau = \tau-\beta/2] \ll \beta$, $\Delta$ the fermionic scaling dimension and $\varepsilon$ the spectral asymmetry~\cite{Sachdev:2015dpa}.
It is known that $\Delta=1/4$, found in the SKY NFL metal, translates to $T$-linear behavior of the resistivity~\cite{Song2017,Patel2018,Chowdhury2018}, while the high-$T$ atomic state and the FL metallic state exhibit $\Delta=0$ and $\Delta=1/2$, respectively. 

\begin{figure}[t]
	\centering
	\includegraphics[width=0.45\textwidth]{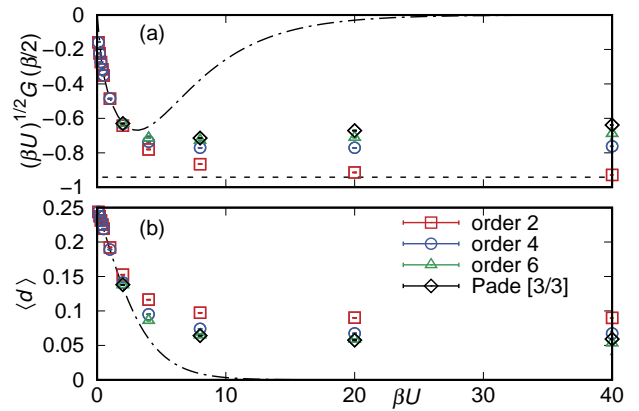}
	\caption{
	        Results of the bold-line diagrammatic theory for different observables of the HA as functions of $\beta U$: 
		(a) $G(\beta/2)$ multiplied by $(\beta U)^{1/2}$, (b) double occupancy.  
		The exact and asymptotic SYK saddle point solutions are presented by the dash-dotted and dashed line, respectively.
	}
	\label{fig:Gtb2_docc_ha_odep}
\end{figure}

We first demonstrate that NFL behavior of the SYK-type is displayed by the full bold-line diagrammatic solution of the HA beyond the second order. 
Although the NFL solution is unphysical \textit{a priori} for the low-temperature HA, which is a Mott insulator, this exercise sets the stage for the non-trivial case of the $2d$ Hubbard model. We compute $\Sigma^{(n)}$ by DiagMC up to $n=6$, which is sufficient for reaching convergence with diagram order. To improve convergence of the iterative solution of Eq.~(\ref{eqn:Dyson}), we use a weighted averaging that mixes in $G^{(n)}$ from previous iterations~\cite{Prokofev2007}. Figure~\ref{fig:scGt} shows the resulting $G(\tau)$ for the half-filled HA. 
As we increase the dimensionless parameter $\beta U$ ($\beta$ is the inverse temperature), $G(\tau)$ crosses over from the high temperature atomic state to a NFL with the SYK long-time scaling.
To more clearly demonstrate the SYK-type character, we plot in Fig.~\ref{fig:Gtb2_docc_ha_odep}(a) the quantity $(\beta U)^{1/2} G(\beta/2)$, which is a constant in the SYK solution (\ref{eqn:sc}) at low enough $T$ (dashed line), but is suppressed exponentially with $\beta U$ in the HA (dash-dotted line) since $-\beta G(\beta/2)$ corresponds to the spectral function near the Fermi level at low temperatures~\cite{Trivedi1995,Gull2008}. The diagrammatic solution matches the exact solution at higher $T$ and crosses over to an unphysical solution with the SYK scaling at lower $T$.
The double occupancies [Fig.~\ref{fig:Gtb2_docc_ha_odep}(b)] of the diagrammatic solutions also saturate at nonzero values for $\beta U \gg 1$, which implies an ever growing (with $U$) potential energy. This is in contrast to the exponentially vanishing double occupancy in the exact physical solution of the HA.

\begin{figure}[t]
	\centering
	\includegraphics[width=0.45\textwidth]{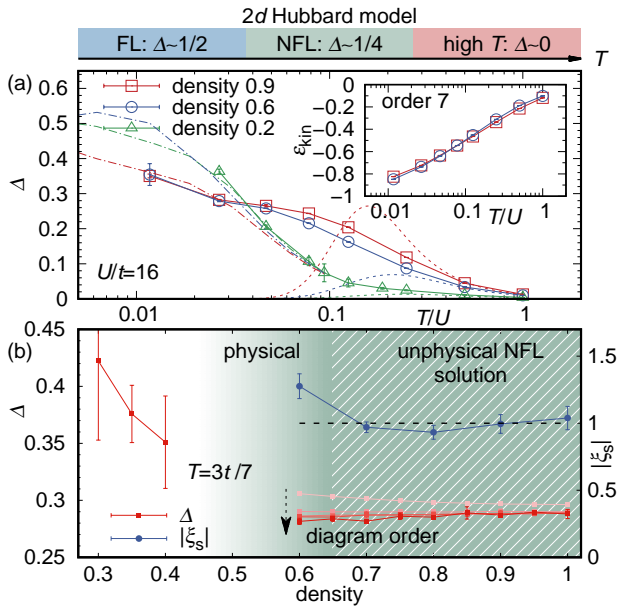}
	\caption{
		Scaling exponent $\Delta$ in the long-time asymptotics of the local Green's function, Eq.~(\ref{eqn:Gansatz}), obtained by the bold-line diagrammatic technique (\ref{eqn:Dyson}), (\ref{eqn:series}) for the $2d$ Hubbard model.
		Panel (a) presents $\Delta$ as a function of $T/U$ on a logarithmic scale at $U/t=16$.  
		The corresponding dashed and dot-dashed lines show the atomic and noninteracting results, respectively.
		The inset of panel (a) illustrates the $\log T$ scaling of the kinetic energy in the intermediate temperature regime. 
		Panel (b) shows $\Delta$ (red symbols) and the location of the singularity (blue symbols) for $T=3t/7$.
		Depinning of the singularity from 
		$|\xi_s|=1$ 
		is observed around density $\approx 0.6$, while $\Delta$ remains around $0.28$.
		Near half filling, the red symbols with color gradient show the order dependence of $\Delta$.
		At intermediate densities the bold diagrammatic series cannot be converged within order 7.
	}
	\label{fig:Gtlong}
\end{figure}


We now turn to the $2d$ Hubbard model, where the bad metal behavior is generically expected. We obtain the full Green's function $G_\mathbf{k}(\tau)$ by the diagrammatic technique (\ref{eqn:Dyson}), (\ref{eqn:series}) in the $n \to \infty$ limit in regimes of parameters where convergence of the series up to $n=7$ can be achieved. 
To 
simplify the analysis and eliminate irrelevant singularities with $\mathrm{Re}\xi <0$, we base the expansion on the homotopic action~\cite{kim2020homotopic} produced by a transformation of the expansion parameter $\xi$ by a conformal map~\footnote{In the following, $\xi$ stands for the transformed expansion parameter for simplicity.}, as illustrated in Ref.~\cite{kim2020homotopic}. To analyze the low-energy behavior of the solution, we fit $G_{\mathrm{loc}}(\tau)=\int_{}^{}\frac{d\mathbf{k}}{4\pi^2}~G_\mathbf{k}(\tau)$ to its long-time asymptotic form~(\ref{eqn:Gansatz}). 

In contrast to the diagrammatic solution of the HA, which exhibits a single crossover from the atomic ($\Delta=0$) to the NFL ($\Delta \sim 1/4$) regimes, two characteristic temperature scales emerge in the vicinity of half-filling. 
As $T$ is lowered, the atomic-state behavior ($\Delta=0$) gives way to an SYK-type NFL ($\Delta \sim 1/4$), which is observed in an appreciable $T$-range, before crossing over to the FL ($\Delta \sim 1/2$) regime. 
Figure~\ref{fig:Gtlong}(a) shows the corresponding $\Delta$ against $T/U$ (at fixed $U/t=16$).
A 
plateau with $\Delta \approx 0.28$ is observed at intermediate $T$ for densities $0.6\lesssim \langle n \rangle \leq 1$, with the temperature range of the NFL state decreasing with doping.
Such a two-step crossover with the NFL as the intermediate-temperature state is generically expected. 
It has been reported not only in the translationally invariant form of the SYK model~\cite{Chowdhury2018} but also in multi-orbital Hubbard models in the context of spin-freezing crossovers~\cite{Werner2008,Haule2009,DeMedici2011,Georges2013,Stadler2015,Hoshino2015,Werner2016,Kim2017,Werner2018}.

The non-local components of the Green's function gradually develop as $T$ is decreased. 
Notably, the kinetic energy, given by the equal-time Green's function between neighboring sites, shows different temperature scalings in the three regimes. 
In the intermediate NFL regime, it turns out to be proportional to $\mathrm{log} T$ (see inset of Fig.~\ref{fig:Gtlong}(a)), in contrast to the $T^2$ Fermi-liquid behavior at low $T$ and the $1/T$ atomic behavior at high $T$. 

The central problem in light of possible misleading convergence is proving the validity of the diagrammatic solution.
To this end, we examine the singularity structure of the self-energy series (\ref{eqn:series}), evaluated at the converged solution $G$, for $|\xi_s|>1$ (physical solution) or $|\xi_s|=1$ (misleading convergence). 
Specifically, we obtain $\xi_s$ using Pad\'e-based approaches~\cite{Hunter1979} with controlled error bars~\cite{Simkovic2019} for the series of the momentum-averaged (local) self-energy at the lowest Matsubara frequency. The results are shown in Figs.~\ref{fig:singularitymap}(c) and (d). 
At density $0.9$ [Fig.~\ref{fig:singularitymap}(c)], as the solution crosses over from the atomic state to the NFL with cooling, $\xi_s$ approaches the unit circle.
For $T/U\lesssim 0.16$, the singularity remains pinned at $\xi_s=1$, which reveals the unphysical character of the NFL solution near half filling. 
However, at a fixed $T=3t/7$ [Fig.~\ref{fig:singularitymap}(d) and Fig.~\ref{fig:Gtlong}(b)], the singularity moves away from the $|\xi_s|=1$ boundary as the doping level increases.

Most importantly, at density $\approx 0.6$, we can detect unpinning of the singularity from the $|\xi_s|=1$ boundary beyond the error bars.
At this doping, the solution retains its NFL character, as seen from Fig.~\ref{fig:Gtlong}(b), where $\Delta$ is plotted alongside $|\xi_s|$ against density. 
Therefore, our data suggest that a physical NFL state is realized in the $2d$ Hubbard model at strong interactions at least near and above 40\% doping.
This finding is qualitatively consistent with the phenomenological phase diagram of cuprates, where the strange metal state occurs in a wide doping regime at $T$ above the superconducting dome.

In summary, we have shown that the numerically exact bold-line diagrammatic solution of the $2d$ Hubbard model exhibits the SYK-type NFL behavior in a broad range of temperatures and doping. 
The physical NFL state is realized at least near and above 40\% doping at $T = 3t/7$, but the physics at larger densities is inaccessible by the bold-line diagrammatics due to its misleading convergence caused by the multivaluedness of the LWF. It is expected that, by continuity, the NFL regime should extend to smaller doping as well, while the spin-freezing analogy~\cite{Werner2018} suggests that it will move closer to the optimal doping with cooling. These results are consistent with the emerging understanding~\cite{Reitner2020} that the regime of strong fluctuations, at least in the charge channel, is fundamentally located on the strongly-correlated branch of the LWF, requiring the transition from the weakly-correlated branch by its mathematical structure. Our findings thus imply that the NFL regime, being intrinsically a transitional state, extends into the weakly-correlated branch across the branching point. 
The role of the branching point in the nature of the NFL state and its possible connection to the pseudo-gap line and reported bad-metal-to-FL transitions \cite{Sordi2012} is an interesting topic for future research.
The unphysical bold-line diagrammatic solution at smaller doping, exhibiting non-trivial and expected physical properties without detectable pathologies, is a remarkable example of a controlled and consistent but delusory theory and a warning for future studies. 

\textit{Acknowledgement}- A.J.K and E.K. are grateful to the Precision Many-Body Group at UMass Amherst, where a part of this work was carried out, for hospitality. This work was supported by EPSRC through grant EP/P003052/1 and by the Simons Collaboration on the Many-Electron Problem. 
We are grateful to the UK Materials and Molecular Modelling Hub for computational resources, which is partially funded by EPSRC (EP/P020194/1).

\bibliography{ref}
\end{document}